# Logic Design for On-Chip Test Clock Generation – Implementation Details and Impact on Delay Test Quality

Matthias Beck, Olivier Barondeau,
Martin Kaibel, Frank Poehl

Infineon Technologies AG
Balanstrasse 73
81541 Munich, Germany

Xijiang Lin, Ron Press

Mentor Graphics Corporation
8005 S.W. Boeckman Road
Wilsonville, OR 97070, US

**Abstract**
*This paper addresses delay test for SOC devices with high frequency clock domains. A logic design for on-chip high-speed clock generation, implemented to avoid expensive test equipment, is described in detail. Techniques for on-chip clock generation, meant to reduce test vector count and to increase test quality, are discussed. ATPG results for the proposed techniques are given.*

## 1 Introduction

Structural testing of digital integrated circuits already has a long tradition in the semiconductor industry. Scan test in combination with automatic test pattern generation (ATPG) for stuck-at and $I_{DDQ}$ test have been the industry standard for many years. Scan based ATPG for structural at-speed test, or delay test, is also well known for about 20 years based on the gate delay fault model [1], path delay fault model [2], and transition fault model [3].

Many publications show that at-speed test is able to detect production defects that are not covered by static tests, e.g. [4]-[9]. Functional at-speed tests require a significant effort to develop compared to delay tests that are automatically generated by ATPG tools. Moreover, delay tests have a known fault coverage because of the usage of ATPG tools. Fault grading is difficult to impossible for functional at-speed tests. Thus, the fault coverage for functional at-speed tests usually remains unknown.

Even though the advantages are multifold and well known, delay test is still far from being as popular as stuck-at fault test is for SOC (system-on-a-chip) devices. This is due to several challenges that must be addressed for delay test to be implemented as part of the overall production test strategy. The first observation is that the number of patterns needed to achieve a high fault coverage for transition fault testing is several times larger than the corresponding stuck-at pattern count. This implies, of course, the same increase in production test time and in vector memory usage on automatic test equipment (ATE). In addition, the transition fault coverage is usually noticeably lower than the corresponding stuck-at fault coverage, [8], [10]-[13]. For path delay testing, a small subset of structural paths is typically selected to be targeted by an ATPG tool. Thus, path delay fault coverage has only a very limited meaning with respect to the overall chip area.

Besides the issues with fault coverage and pattern count, special design measures are required to run at-speed tests for several hundred MHz designs on ATE with significantly lower speed capabilities. If high frequencies are generated on-chip for delay testing, then the ATPG tools need the ability to control the clock generation mechanism for all clock domains. In addition, the at-speed test data needs to be synchronized to the external low-speed tester.

Looking at the still incomplete and growing list of issues, it is fortunate that solutions have improved over time and are continuing to improve. With the latest developments in the area of pattern compression, the impact of increased pattern count is significantly reduced [14]-[17]. As design rules for delay test are better understood and ATPG tools become more and more mature to solve complex testability problems, fault coverage is also increasing. Design engineers select paths for path delay test more carefully in a manner that test results can be better correlated to production problems. Path delay tests can even be used for speed grading [12].

Several high frequency designs, that are tested at-speed with a low frequency ATE, have been reported over the past few years. High-speed clock signals are generated on-chip using the internal functional PLL. Examples for high-speed on-chip clock generation, but no implementation details, can be found in [10], [11], [13], and [18]. With this approach, test data is synchronized to the external low-speed tester using a two-clock source strategy. Scan test data is shifted using a slow clock signal generated by the ATE. The required at-speed delay test pulses are generated on-chip by the PLL.

This paper presents a delay test concept for SOC devices with multiple clock domains. Circuitry for high-speed on-chip clock generation, implemented to avoid expensive test equipment, is described in detail. To the authors' knowledge, this is the first publication that presents implementation details for high-speed on-chip test clock generation. Enhancements to the presented clock generation circuitry, that are meant to improve fault coverage and pattern count, are discussed. Experimental ATPG results prove that the capabilities of the on-chip clock generation circuitry have an impact on achievable fault coverage and pattern count.

The paper is organized as follows. The next section explains delay test fundamentals for SOC designs. Section three addresses implementation details of the on-chip clock generation circuitry. The fourth section discusses the delay test generation tasks from the ATPG tool perspective. Section five describes the experiments and the results achieved, respectively. Conclusions and an outlook are presented at the end of the paper.

## 2 Delay Test for State-of-the-Art SOC devices

Delay test is a two-vector (cycle) test. The first vector initializes the test. The second vector launches the desired transition at the source of a path under test and propagates the launched transition to an observation point. Test responses are captured with the next active clock edge at the observation point. Thus, test frequency requirements between *initial* and *launch* vector are relaxed, while the clock between *launch* and *capture* has to be operated *at* the functional speed of the device under test (DUT).

Two techniques are known to apply two-vector tests using a standard scan architecture, [19], [20]. *"Scan shifting"* or *"launch-from-shift"* shifts the initial vector into the scan chains and generates the launch vector by using the last shifting cycle. In *"functional justification"*, *"broadside"*, or *"launch-from-capture"* the initial vector is shifted into the scan chains. The launch vector is the functional response of the DUT to the initial vector. *"Scan shifting"* requires the ability to switch the scan-enable signal at-speed. In addition, the risk of testing non-functional paths is higher for *"scan shifting"* since both vectors are completely independent of the DUT function. Due to the disadvantages of *"scan shifting"*, it was decided to implement delay testing based on *"junctional justification"*. The remainder of the paper refers to this delay test technique.

SOC devices implement multiple functional clock domains. An at-speed test has to test all clock domains at their own functional speeds. Testing one clock domain after another is a potential solution for multi-clock domain delay test. However, the logic between the clock domains remains untested and the number of patterns is comparably high. Testing all clock domains simultaneously requires separate clock pins for each domain and a complex clocking procedure if an ATE is used to apply the test clock [8]. As already explained in the introduction, the usage of an ATE for clocking also requires a high speed clock generator to apply at-speed clock pulses for launch and capture. A high-speed clock generator is usually not available in low-cost ATE.

Even if a high-speed clock generator is available, it is doubtful that a high-speed clock can be transmitted into the DUT. Today's SOCs, especially micro-controller devices, communicate to their environment at a comparably low speed. Therefore, the device pads are slow compared to the internal frequency. The application of an external high-speed clock would require the additional implementation of a high-speed pad that is capable of transmitting the clock signal.

To overcome these issues the functional on-chip PLL can be used to generate at-speed clock pulses for test purposes. Figure 1 shows a device with two functional clock domains. The PLL generates two independent high-speed clock signals, *pll_clk_1*, *pll_clk_2*, for the two clock domains, derived from the slow external clock *ext_clk*. The delay test clocking principle is shown in Figure 2. Delay test data is shifted using the slow external *scan-clk*. After shift has finished, *scan-en* is switched off and two at-speed clock pulses are filtered out of *pll_clk_1* and *pll_clk_2*. *clk_1* is derived from *scan-clk* and *pll_clk_1* using the *clock pulse filter* I (CPF1) which is controlled by *scan-en*. A similar mechanism is implemented for *clk_2*.

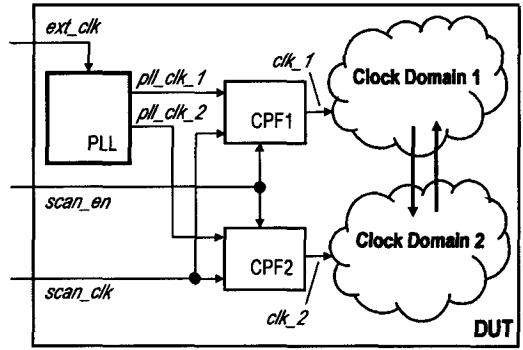

Fig. 1: Device with clock pulse filters (CPF) for each of the two clock domains

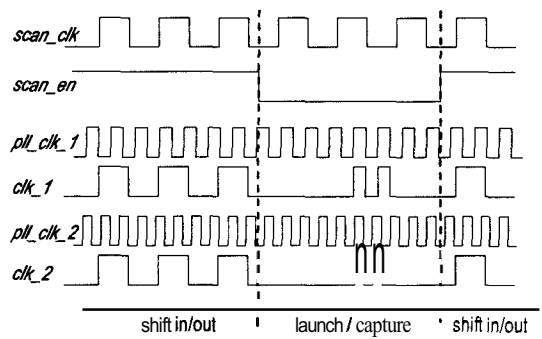

Fig. 2: Delay test clock for two clock domains

The clock pulse filters replace the clock multiplexers that are used to switch between *scan-clk* and functional clock for standard low-speed stuck-at scan test. Implementation details of the CPF and the exact clocking scheme are explained next.

## 3 On-Chip High Speed Clock Generation

Figure 3 shows the schematic of the CPF. The CPF is an add-on block to the PLL. Input signals are *pll_clk*, *scan-clk*, and *scan-en*. The output signal *clk-out* is applied to the logic.

Figure 4 shows the corresponding waveform diagram. While *scan-en* is '1', *scan-clk* is directly connected to *clk-out*. When *scan-en* is set to '0' the output of the clock gating cell (CGC), *cgc_clk_out*, is connected to *clk-out*. The implementation of CGC makes sure that no glitches or spikes appear on *clk-out*. CGC is enabled by the signal *hs_clk_en* that is generated from the five-bit shift register. The shift register is clocked by *pll_clk*. According to Figure 4, a single *scan-clk* pulse is applied after *scan-en* is set to '0'. This clock pulse generates a '1' that is latched by the flip-flop and shifted through the shift register.

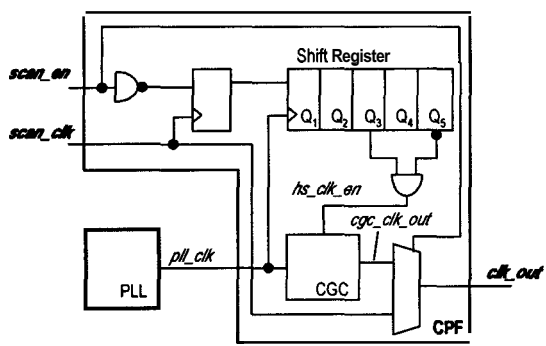

Fig. 3: Schematic of the clock pulse filter

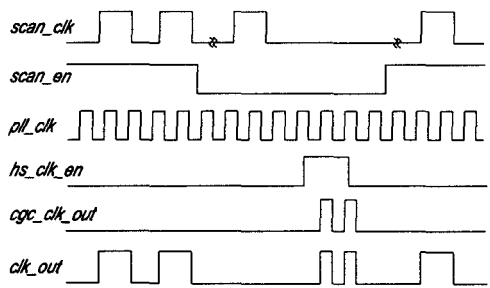

Fig. 4: Clock pulse filter (CPF) signals

After three *pll_clk* cycles, *hs_clk_en* is asserted for the next two *pll_clk* cycles. As CGC is enabled during that period, exactly two PLL clock pulses are transmitted from the PLL to *clk-out*. Additional logic, not shown in Figure 3, ensures that the CGC is always enabled in functional mode.

The area required to implement the CPF is negligible. The entire CPF consists of ten standard digital logic gates per clock domain only, as shown in Figure 3. The CPF introduces an additional delay on the clock-tree between PLL and functional logic. This delay is compensated during clock-tree balancing. It has therefore no impact on the chip timing.

There is no need for a high-speed relation between *scan-clk* and *scan-en*. After the test data is shifted into the scan chains, *scan-clk* is stopped and *scan-en* is set to '0' with relaxed timing. Once *scan-en* is stable, *scan-clk* is clocked one time to trigger the enable mechanism that releases the two high-speed launch and capture clock pulses. *scan-en* is asserted and *scan-clk* operation is continued after the device is stable again. There is no need to synchronize the internal PLL clock to *scan-clk* or *scan-en*. The technique requires, of course, that a PLL clock signal is permanently available during the entire delay test.

Besides the advantages already mentioned, the implementation is also testing the entire functional clock generation circuitry during delay test. Only scan shifting is performed with relaxed speed. Furthermore, since the scan chains are not used in functional mode there is no need to test them at-speed.

## 4 Delay Test ATPG

Delay test ATPG has to consider at least two consecutive clock cycles, as delay test is a two-vector test. Due to the sequential nature of the ATPG problem, delay test ATPG is more complex than stuck-at ATPG. The use of more than one clock cycle during ATPG is already known for stuck-at ATPG. Additional clock cycles are used to initialize sequential elements such as non-scan cells and RAMs – known as "clock *sequential*" and **"RAM sequential"** ATPG, respectively. This adds even more complexity to the delay fault ATPG problem. But especially access to RAMs is important for delay testing as critical functional timing paths frequently involve functional paths from and to RAMs.

Delay test is an at-speed test in which all clock domains could run at their functional speeds simultaneously if tested in parallel. To prevent wrong capture values for signals passing between clock domain boundaries, it is important that the ATPG tool is able to understand complex clocking schemes [21].

If at-speed clock pulses are generated by an on-chip PLL, then the ATPG tool must be able to control the on-chip clock generation. If the CPF from Figure 3 is used, clock generation is controlled by *scan–en* and *scan–clk* only. To generate "clock *sequential*" and "*RAMsequential*" patterns an enhanced CPF as well as a more advanced control mechanism is required. This is not an issue when an external clock is used since the ATPG tool can decide how many clock cycles need to be applied. To get the same freedom with on-chip clock generation, a dedicated control protocol to setup the PLL from the ATPG tool is required.

The efficiency of an ATPG tool is significantly reduced if every cycle into and through the PLL CPF needs to be simulated. Six or more pulses of *pll_clk* during functional mode may be required to produce a desired clock pulse pair at *clk_out*. Therefore, procedures referred to as *named capture procedures* were developed. *Named capture procedures* provide a simple behavioral model of the clock generation logic. The model describes the behavior of the output clock signal *(elk-out)* in response to primary input signals *(scan–en, scan–clk)*. As a result, the *named capture procedures* can model internal clock generation logic as a couple of internal clock pulses during ATPG. When the patterns are saved for ATE, the internal clock pulses (*clk_out*) are converted to the corresponding primary input signals that will produce them.

An additional consideration when creating delay patterns is that the device functional IOs usually cannot perform at the high frequencies of the PLL clocking. In addition, it is difficult at the ATE to reference the external IO timing to an on-chip generated clock. Therefore, any inputs driven by the ATE or output pin responses must not require at-speed timing in relation to the clocks or other signals. As a result, it is normally necessary to mask outputs from being observed and to ensure that primary inputs only change well ahead of the launch clock pulse. *Named capture procedures* support this by only forcing inputs and measuring outputs when they are explicitly specified.

Utilizing CPF logic for delay testing can be used for transition fault model and path delay fault model patterns. It can also be extended to provide clocking when applying memory tests through the scan logic. This technique is sometimes referred to as macro testing and enables at-speed testing of memory operations without adding any memory test logic [22].

## 5 Case Study

To determine the impact of on-chip clock generation on fault coverage and pattern count, ATPG experiments based on the transition fault model were performed. The path delay fault model was not used for comparison since the path delay fault coverage is highly dependent on the testability of the pre-selected paths. ATPG settings are compatible throughout all experiments to ensure comparability of results.

Experiments were done using the netlist of a 0.13µ technology micro-controller SOC device with a total logic gate count of 1.36M. Two synchronous functional clock domains operate the device, running at 75 MHz and 150 MHz. 357 balanced internal scan chains, based on an EDT architecture, [15], [17], with 36 external scan channels, are implemented for 65.7k multiplexed scan cells. The number of non-collapsed faults considered is 4.3M. This number is identical to the stuck-at fault count as both fault models are targeting two faults at each gate terminal. However, it is important to note that not all fault sites that are testable for stuck-at faults are also testable for delay faults.

### 5.1 Experimental Setup

For all experiments, but (a), the transition fault coverage and pattern count was determined.

(a) *Stuck-at test using single external clock:* Both clock domains are operated by a common external scan clock. RAM sequential pattern are not considered to allow identical ATPG settings for all experiments.

(b) *Transition test using single external clock:* Same setup as in (a). This experiment reveals the maximum achievable transition fault coverage if the device would consist of a single clock domain. The scan clock and IOs are controllable and observable from the ATE.

(c) *Transition test using on-chip clock generation:* For each of the two functional clock domains a CPF block as shown in Figure 3 is used, providing two clock pulses. Additional constraints required for the on-chip clock generation and ATE execution are added (no launch or capture using test controller clock or system reset, scan enable always inactive during capture phase, IOs not used, feedback paths through bidirectional pads not utilized).

(d) *Enhanced CPF block to generate up to 4 high-speed clock pulses and inter-domain test:* The same experiment as in (c), but the CPF blocks are enhanced and able to provide two, three or four clock pulses. In addition, the CPF blocks provide the capability to generate tests for inter-domain signals crossing the boundaries of the synchronous clock domains. These tests apply a launch pulse in one clock domain and a capture pulse in the other clock domain.

(e) *Single external* clock *with constraints and IO masking:* This experiment gives the coverage that would be reachable by the most flexible CPF blocks possible. Instead of the two CPF blocks, a single external clock is used as in experiment (b). All other constraints required by an internal clock generation are still taken into account as introduced in experiment (c).

## 5.2 Experimental Results

Test generation results for both the stuck-at faults and the transition faults are shown in Table 1. Column *"TC"* depicts the test coverage achieved in percentage while column *"# Pattern"* gives the associated pattern count. Tool runtimes are not reported because experiments were run on different type of CPUs. Tool runtimes were in the range of one day for stuck-at faults and about one week for transition faults.

The test coverage reaches **98.7%** for stuck-at faults. The remaining stuck-at faults are, 1% ATPG untestable, 0.3% aborted. The ATPG efficiency for transition fault experiments (b), (c), and (d) is also above **99%.** For experiment (e) the ATPG efficiency is **97.7%.**

Comparing results (a) and (b) the coverage for transition faults is already **3.7%** lower than the stuck-at fault coverage, even if multiple clock domains and high-speed on-chip clock generation are not considered.

Experiment (c) shows that the coverage for a real applicable transition test using on-chip clock generation is much lower compared to the reference value of experiment (b). Main reason for the coverage drop are additional test generation constraints required to enable the use of CPF blocks as described in Figure **3.** Moreover, external clock generation allows multiple pulses to initialize non-scan cells and RAMs while only exactly two clock pulses are generated by the internal CPF block. Another difference from experiment (b) is that both clock domains are tested with independent clocks and only one CPF block is used per scan load. Therefore, no inter-domain test is possible.

To overcome the coverage reduction caused by the simple CPF used in experiment (c) a more flexible CPF is assumed in experiment (d). The new capabilities of the enhanced CPF also enable the generation of three or four clock pulses and allow the launch and capture between the different clock domains. The enhanced capabilities of the CPF increase the coverage by 0.6%.

Experiment (e) again uses a common external clock for both clock domains. This experiment is used to determine the coverage that would be achievable for a most flexible CPF. It also excludes **any** effects that might be caused by the interaction of the two clock domains or limitations of the ATPG tool dealing with on-chip clock generation. The reached coverage is still 6.6% lower than the reference value of experiment (b). Since the only differences between experiments (b) and (e) are the constraints added to make the test applicable at the ATE, these limitations and restrictions must be the reason for the coverage reduction.

| Fault Type | Exp. | TC [%] | # Pattern |
|---|---|---|---|
| Stuck-at | (a) | 98.68 | 6,464 |
| Transition | (b) | 94.96 | 29,147 |
| | (c) | 87.38 | 68,484 |
| | (d) | 87.99 | 69,785 |
| | (e) | 88.38 | 58,351 |

Table 1: Experimental Results

The pattern count for experiment (b) is nearly five times higher than the stuck-at pattern count. Using on-chip clock generation, experiment (c) and (d), increases the pattern count again by more than a factor of two. Experiment (e) shows that an enhanced CPF reduces the pattern count by more than 15% compared to experiment (d).

## 6 Conclusions

Even though delay defects become increasingly important for new technologies, high-speed testers are hardly affordable for SOC devices targeting consumer products. Therefore, high-speed on-chip clock generation for delay testing is one cornerstone of future SOC test strategies.

For the device under investigation the transition test coverage is reduced by more than **7%** when simple two-pulse on-chip clock generation is implemented. Moreover, increased pattern count requires a more extensive use of an on-chip technique to reduce scan chain length. Only using this technique the observed pattern count can be loaded into the ATE vector memory without truncation of the test pattern set.

Experiments have shown that enhancements in on-chip clock generation capabilities improve the test coverage and reduce the pattern count. To achieve further improvements future work will focus on two main objectives.

ATPG development will concentrate on enhanced fault classification and fault grouping. Many faults included in the transition fault coverage report are actually non-functional faults and will make the coverage appear lower than the actual quality of the test. An attempt will be made to classify and group these faults as non-functional scan path, low-speed IO, and other faults that cannot cause the device to fail at-speed operation.

Circuit development will concentrate on further analysis of root causes for design related coverage reduction. For those faults not tested but relevant for product quality, counter measures need to be developed to achieve an acceptable coverage. For example, at-speed testing of logic between clock domains has been avoided in the past. The experiments show that these tests, that are meaningful for signals between synchronous clock domains, improve the coverage at least to some extent.

In the future it needs also to be proven that enhanced delay tests guarantee the same quality level as today's functional at-speed tests.

**7** Acknowledgement

The authors would like to acknowledge the contributions of Eric Guan of Infineon Technologies Singapore who implemented the first version of the CPF module.

Part of this work has been funded within the AZTEKE project under label 01M3063A by the German ministry for education and research (BMBF).